\documentclass[aps, amsmath,amssymb, amsfonts, nofootinbib, notitlepage, twocolumn, preprintnumbers]{revtex4-1}
\usepackage{graphicx}
\usepackage{dcolumn}
\usepackage{bm}
\usepackage{bbold}

\graphicspath{ {./Desktop/} }

\begin{document}

\preprint{APS/123-QED}

\title{Nucleon-Antinucleon Annhiliation at Large-$N_c$}

\author{Thomas D. Cohen}
 \email{cohen@physics.umd.edu}
\author{Brian McPeak}
 \email{bmcpeak@umd.edu}
 \affiliation{%
Department of Physics and Maryland Center for Fundamental Physics\\University of Maryland, College Park, Maryland\\
}%
\author{Bendeguz Offertaler}
 \email{bendeguzoffteraler@gmail.com}
\affiliation{%
Mongtomery Blair High School, Silver Spring, Maryland\\
}%

\date{\today}

\begin{abstract}
Nucleon-antinucleon annihilation in the large $N_c$ limit  of QCD in the Witten regime of fixed velocity is considered with a focus on the spin and isospin dependence of the annihilation cross-section. In general, time-reversal and isospin invariance restricts the annihilation cross-section to depend on 6 independent energy-dependent terms.  At large $N_c$, a spin-flavor symmetry emerges in the theory that acts to further restrict the dependence of the annihilation cross-section to three of these terms;  the other terms amount to  $1/N_c$ corrections.  Assuming dominance of the leading order terms, several identities are derived that relate annihilation in different spin-isospin channels. A key prediction is that for unpolarized nucleons in Witten kinematics, the proton-antiproton annihilation cross-section should be equal to the proton-antineutron annihilation cross-section up to corrections of relative order $1/N_c$. Unpolarized nucleon-antinucleon annihilation data appears to be consistent with this expectation.  
\end{abstract}

\maketitle

\newcommand{\updownarrows}{\mathbin\uparrow\hspace{-.5em}\downarrow}
\newcommand{\downuparrows}{\mathbin\downarrow\hspace{-.5em}\uparrow}

\section{Introduction}

Quantum chromodynamics (QCD) has long been accepted as the underlying theory for hadronic physics. However explicit calculations for many phenomena in the theory remain elusive.  On the one hand, perturbation theory in the coupling fails at small momenta.  On the other hand, Monte Carlo evaluation of the Feynman path integral of a Euclidean space lattice version of QCD is currently the only viable non-perturbative method currently available to obtain observables directly from the QCD Lagranagian and it has significant limitations.  Lattice techniques are very well suited for static properties of hadrons. It is also possible to study low energy scattering of hadrons using  L\"uscher's  method\cite{Luscher1,Luscher2,Luscher3,Luscher4} of relating S-matrix information such as phase shifts to energy levels in a finite spatial box. While there has been considerable recent progress in developing this method into a practical tool, it is clear that this approach becomes increasing unwieldy as energies increase. It is clear that the L\"uscher approach is not viable for scattering observables associated with higher energy scattering for which phase space allows a very large number of  particles in the final state.

While a direct calculation of high-energy scattering observables directly from QCD is well beyond the present state of the art, it is possible to learn {\it something} about some  high-energy scattering observables from the large $N_c$ limit of QCD.  Recall that 't Hooft\cite{hooft} recognized  long ago that $1/N_c$,  where $N_c$ is the number of colors, can be used as an expansion parameter in QCD.  The underlying premise of this approach is that a world with an infinite number of colors is qualitatively similar to the world of $N_c=3$. Given this premise one should replace the $SU(3)$ gauge group with $SU(N_c)$. This makes it possible to study observables of interest in the limit where $N_c$ goes to infinity and to calculate systematic corrections in a power series in $1/N_c$.  The difficulty with such an approach is that the large $N_c$ limit of QCD, while providing great simplifications compared to QCD at $N_c=3$, in general is still not tractable by analytic means except in special cases where gluons do not play a role as dynamical degrees of freedom, such as the `t Hooft model in which  space-time is 1+1 dimensional \cite{hooft2} or QCD in the regime in which quark masses are much larger than the QCD scale\cite{witten,tdc1}.  Fortunately, even when one cannot directly compute even the leading order behavior, it is often possible to deduce some properties. For example, it is often possible to deduce the $N_c$ scaling behavior of various observables which may allow for qualitative predictions of the relative size of various observables---under the assumption that  three can be considered as large for the purposes of the analysis.

It turns out that among the phenomena for which predictions at large $N_c$ exist are certain properties of scattering observables in nucleon-nucleon scattering\cite{cohen1,cohen2,cohen3}. All of these depend on the system being in the regime of Witten kinematics\cite{witten} in which the incident momentum, along with the nucleon mass, scales as $N_c^1$---ensuring that the velocity is held fixed as $N_c$ diverges.  Note that  Witten kinematics is intrinsically a high energy regime in that as $N_c$ goes to infinity, the initial kinetic energy grows with $N_c$ and the number of mesons kinematically allowed to be created in the scattering process also grows with $N_c$. A key result is that spin-flavor dependence of a certain class of scattering observables\cite{cohen1} is constrained by an emergent symmetry at large $N_c$\cite{SF1,SF2,SF3,SF4,SF5,SF6,SF7,SF8}. While the large $N_c$ analysis of the spin-flavor dependence of nucleon-nucleon scattering is more than a decade old\cite{cohen1}, it has basically not been empirically tested against real world data in order to determine the extent to which the large $N_c$  limits are a reasonable caricature of the physical $N_c=3$ world with regard to scattering.   There are a couple of reasons for this: first, to get  the full spin-flavor dependencies one needs to have access to data for both proton-proton and proton-neutron scattering with polarized beams and targets. Such data are relatively rare. Moreover, experimentalists in the field do not typically analyze their data in terms of the types of observable for which the large $N_c$ analysis applies.  
 
There has been one attempt to analyze real world data\cite{cohen4} to see if the predicted spin-flavor dependence due to large $N_c$ can be seen at least approximately. That attempt exploited the fact that formally, the large $N_c$ analysis should apply when $N_c$ is large enough to justify a semi-classical treatment, which occurs when the incident momenta are much larger than the typical scales of QCD. It is easy to see that at a formal level this condition can be met for sufficiently large $N_c$, even below the elastic threshold which occurs for momenta $p \sim  \sqrt{M_n m_\pi} \sim N_c^{\frac{1}{2}} $. This implies that if $N_c$ were large enough, the predicted spin-flavor dependence should be evident in elastic scattering just below the elastic threshold.  Empirically, the predicted pattern is not seen in the elastic scattering data even approximately\cite{cohen4}. This is hardly surprising. Firstly, in the real world in which $N_c=3$, it may be reasonable in some circumstances to argue that $N_c$ is large, but it stretches credulity to argue that the square root of three should be considered large as well. Secondly, the momentum scale at the elastic threshold is anomalously small due to the approximate chiral symmetry of QCD which leads to a light pion. Thus, the value of $N_c^{1/2}$ needed to make the scattering semi-classical needs to be particularly large to compensate for the small size of $m_\pi$.

In this paper we extend this large $N_c$ analysis to the problem of nucleon-antinucleon scattering at large $N_c$ with an emphasis on the  annihilation cross-section. In particular, we determine the leading order spin and flavor dependence of the annihilation cross-section. It is interesting to show that the kind of analysis used in nucleon-nucleon scattering is generalizable to a new class of problems. More significantly, the annihilation cross-section is something that experimentalists conventionally measure and so it provides a possible testing ground for the ideas underlying  scattering in large $N_c$ QCD.

 \section{Baryons at large $N_c$}

The present analysis of nucleon-antinucleon annihilation is largely analogous to previous work on nucleon-nucleon scattering.  It is therefore useful to review baryon-baryon scattering at large $N_c$ and the analysis of single baryons upon which it depends before turning to the problem at hand. This section briefly reviews baryon properties at large $N_c$ and the following one reviews baryon-baryon scattering. The basic framework for treating both of these problems  was largely laid out in Witten's classic 1979 paper\cite{witten}. In that paper, intuition about baryons in large $N_c$ QCD is developed in the limit where all quarks are heavy: $m_Q \gg \Lambda_{\rm QCD}$. 

A critical first piece of this analysis is the argument that generic baryon properties  can be accurately described by a mean-field treatment. It was then argued that $N_c$ scaling rules such as the fact that the baryon mass scales with $N_c$ hold even away from the heavy quark limit. A key observation of Witten was that properties of baryons scale with $N_c$ in precisely the same way that properties of solitons scale with $1/g^2$ where $g$ is the coupling constant. This leads to the natural expectation that generic properties with scaling in soliton models such as the Skyrme model \cite{Skyrme,ANW,SkyrmeReview} faithfully reproduce the large $N_c$  scaling behavior of QCD, with a semi-classical treatment of the weakly coupled theory playing the role of the mean-field treatment.  Following Witten's paper it was rapidly realized that there were a number of  predictions of Skyrme type models which appeared to be model independent in the sense that they held regardless of the parameters of the models or the details of the Lagrangian; they only depended on the symmetries of the theory and semi-classical treatment\cite{MI1,MI2,MI3,MI4}. 

A large number of the model-independent predictions of baryons at large $N_c$ concern symmetries. For simplicity, we will restrict our attention in this paper to systems with two degenerate light flavors.  At the level of large $N_c$ QCD for baryons, consistency relations imply the existence of an emergent contracted SU(4) symmetry (a special case of the more general property of $N_f$ degenerate flavors yielding a contracted SU($2N_f$) symmetry)\cite{SF1,SF2,SF3,SF4,SF5,SF6,SF7,SF8}.  The algebra associated with this contracted SU(4)  group is specified in terms of 15 generators, $J_i$, $I_a$ and $X_{ia}$ with $i$ and $a$ running from 1 to 3.  The commutation relations are 
\begin{align}
[J_i,J_j] &=\epsilon_{i j k} J_k  \nonumber\\
[I_a,I_b]& =\epsilon_{a b k} I_k \nonumber\\\
[J_i,X_{j a}] &=\epsilon_{i j k} X_{k a} \\
[I_a,X_{i b}]& =\epsilon_{a b c} X_{i c} \nonumber\\
[J_i,I_a] &=0 \nonumber\\ 
[X_{a i},X_{j b}]&=0\nonumber
\end{align}
The relevant representation of this group is infinite dimensional and consists of states with $I=J$.  

The key thing which enables one to make predictions is that for the nucleon states of interest, the matrix elements of all of the generators are of order unity, that is $N_c^0$. In the case of matrix elements of $J^i$ or $I^a$ this is obvious since the nucleons have spin and isospin 1/2. For the case of $X_{ia}$ it is less obvious but still true. It ultimately follows from the fact that this is a contracted symmetry. In effect $X_{i a} = G_{i a}/N_c$ where $G_{i a} \sim N_c$ is the naturally arising  object if there were an ordinary $SU(4)$ rather than a contracted one. Thus, for example, the matrix elements of $G^{ia}$ would be of order $N_c$ because these elements are $\langle N | \bar{\psi} \gamma^i \gamma_5 \tau^a \psi | N \rangle$. This is the strength of the axial vector coupling, which is of order $N_c$ due to normal large $N_c$ counting rules.

Model-independent relations can then emerge when observables are related by Clebsch-Gordan coefficients of the emergent symmetry. In the Skyrme model the hedgehog structure of a single isolated classical soliton solution breaks both rotational and isospin symmetry. Physically these hedgehogs correspond to superpositions of  physical states with well-defined spin and isospin quantum numbers. Since the results of interest are model-independent, it is sufficient to use the simplest version of the Skyrme model, which uses only pion degrees of freedom.  These are collected in a single matrix-valued field $U(\vec{x},t)$ with $U \in {\rm SU}(2)$. The classical zero modes associated with this breaking become collective coordinates and must be requantized in order to restore the symmetries of the physical states\cite{ANW}. These requantized states are connected by the same group structure as the one in the underlying large $N_c$ theory and hence respect all of the same symmetry relations.

A single hedgehog for a soliton with  baryon number unity may be written as 
\begin{align}U(\vec{x},t) & = A(t) \, U_0 (\vec{x})\, A^{-1}(t) \; \; \; {\rm with} \\  U_0(\vec{x}) &= e^{iF(r)\tau\cdot\hat{x}} \nonumber \end{align}
 where $F$ is a radial function that satisfies $F(0) = \pi, F \rightarrow 0$ as $r \rightarrow \infty$ and minimizes the energy subject to that constraint. $A(t) \in {\rm SU}(2)$ is a spatially uniform matrix which specifies the particular orientation of the hedgehog--that is, the degree to which spin and isospin are aligned.  
$A$ can be written as $A = a_0 + i\bf{a} \cdot \bf{\tau}$ with $a_0^2+{\bf a}^2=1$.  Since  $A(t)$  is specified by  four variables and one constraint,  it can always be specified by three parameters---for example the Euler angles.  These parameters become the collective coordinates for the single baryon problem.     Each of the requantized physical states $\lvert \, m_s \ m_i \, \rangle$ can be interpreted as corresponding to collective wave functions in terms of the $A$. Up to overall normalization constants, these collective wave functions turn out to be the Wigner $D$-matrices \cite{SkyrmeReview}. Physical quantities are obtained from appropriate integrals over $A$ weighted appropriately by the quantity of interest and the collective wave-functions

Let us return to those relations that hold independently of the parameters in the model or the number of terms in the Lagrangian.  Such relations apparently depend entirely on the fact that the underlying baryon has a hedgehog structure and the physical states are obtained using collective wave functions that are given by Wigner $D$-matrices\cite{MI1,MI2,MI3,MI4}. As such, it is highly plausible that they are consequences of the emergent contracted SU(2$N_f$) at large $N_c$. In numerous cases it has been shown that it is indeed the case that such a relation as is seen in the Skyrme model can be derived directly from the group-theoretic structure with no additional input. In addition there also exist many relations in Skyrme type models that involve the behavior of a quantity as the chiral limit of $m_\pi=0$ is approached but are otherwise independent of the parameters of the model.  In all known cases, such behavior can be inferred from large $N_c$ chiral perturbation theory, a variant of chiral perturbation theory that builds in the consequences of the contracted SU(2$N_f$) (including the degeneracy of the nucleon and the $\Delta$)\cite{MI2,MI3,MI4}.  Given this, it is highly plausible that {\it all} examples of relations in Skyrme type models that hold independently of the parameters of the model are,  in fact, true model-independent results that hold for group-theoretical reasons at large $N_c$.  In what follows we shall assume that this remains true for nucleon-nucleon scattering and nucleon-antinucleon as well.

\section{Nucleon-nucleon Scattering}

Nucleon-nucleon scattering at large $N_c$ was also first discussed in Witten's classic 1979 paper\cite{witten}.  Here again the analysis was first done in the context  of the heavy quark limit for simplicity; it was argued that the natural language for describing such processes is time-dependent-mean-field theory (TDMFT).  A key point in this analysis was that in order to have a smooth large $N_c$ limit in TDMFT, the initial conditions must be taken with fixed velocity as $N_c \rightarrow \infty$ rather than with  fixed momentum.   This in turn means that the initial momentum is of order $N_c^1$, as is the initial kinetic energy.  The regime of large $N_c$ with fixed initial velocity is referred to as ``Witten kinematics''.   The analog to TDMFT for situations including light quarks in models such as the Skyrme model---which are expected to accurately encode the leading large $N_c$ scaling behavior and all consequences of emergent large $N_c$ symmetries---is time-dependent classical field theory.

In ref. \cite{witten}, Witten identified  TDMFT as the appropriate formulation of baryon-baryon scattering and also identified the appropriate kinematic regime being momentum of order $N_c^1$.  However, it is striking that ref. \cite{witten} did not identify precisely {\it what} is being calculated in a TDMFT calculation.    It is noteworthy that TDMFT does {\it not} yield S-matrix elements---the basic quantum mechanical objects characterizing the scattering\cite{cohen1}.  Similarly, one cannot directly compute the total cross-section, the total elastic cross-section or the differential elastic cross-section.  Rather, as noted in ref. \cite{cohen1},  TDMFT (or the analogous time-dependent classical solutions in Skyrme type models)  allow one to compute variables associated with flows of conserved quantities such as baryon number or energy\cite{cohen1}.  (Although, it was subsequently shown that by studying the breakdown of the regime of validity of TDMFT as a function of impact parameter, one can get information about the total cross-section and elastic cross-section\cite{cohen2,cohen3})  Using TDMFT, in Witten kinematics one can, for example, meaningfully compute the angular dependence of the outgoing energy relative to the beam axis, by doing TDMFT calculations with different impact parameters and averaging over impact parameters.  In essence, these quantities correspond to averages over many S-matrix elements.  

At first sight,  this seems quite promising: there are physically relevant quantities which can be obtained from mean-field or classical dynamics in Witten kinematics.  Unfortunately, in practice this is of limited utility.  Note that the explicit form of the TDMFT equations derived in ref. \cite{witten} are only valid in the regime in which all quark masses are much larger than $\Lambda_{\rm QCD}$.  For realistic quark masses of relevance to physical nucleons, the form of the TDMFT for QCD equations is unknown.  Of course, one can always use a model such as the Skyrme model in place of QCD.  However, such models are {\it not}  QCD and thus cannot reproduce the detailed results of large $N_c$ QCD.  Such models have an important virtue however. They are believed to faithfully reproduce the leading $N_c$ scaling of observables and correctly encode both any approximate symmetries of QCD (such as chiral symmetry) and the leading order effects of emergent spin-flavor symmetry of baryons.     Thus, one can use models of this sort as a tool to identify results of large $N_c$ QCD which depend only on the symmetry properties and scaling laws.  We note in passing that for certain baryon observables the large $N_c$ and chiral limits are not uniform and results may depend on the order of limits\cite{MI2,MI3,MI4}; the semi-classical analysis of soliton models corresponds to taking the large $N_c$ limit first with fixed quark mass prior to the chiral limit; one can then do a chiral expansion for each coefficient in a $1/N_c$ expansion.  

Topological solitons in chiral field theories  based on hadronic degrees of freedom such as the Skyrme model\cite{Skyrme, ANW,SkyrmeReview} can be used to identify  model-independent properties of scattering as well as static properties. The analysis to do this was developed in \cite{cohen1} and this approach will be briefly reviewed here. It turns out that the primary predictions involve the spin-flavor dependence of various observables. Of course, it is precisely these that we expect to be associated with the contracted SU(2$N_f$) symmetry. Let us see how to use these models to deduce the spin-flavor dependence of variables associated with the collective flow of conserved quantities such as energy, momentum, or baryon number.  One begins the analysis by imagining that one has access to an arbitrarily  large set of solutions of the classical (or mean-field) scattering processes associated with a given initial velocity that fixes the incident energy.  These classical solutions all involve initial conditions of two widely separated rotated hedgehog baryons moving towards each other offset by an impact parameter ${\bf b}$, a two dimensional vector. For concreteness since the experiments we have in mind are typically fixed-target experiments, we can restrict our attention to situations in which one soliton is stationary and the other is moving toward it. Thus, the initial conditions depend on the velocity, the scattering axis and 8 more collective variables---2 for the components of the impact parameter vector and 6 to specify the orientation of the two initial hedgehogs.
 
The classical solutions are, in effect, movies. The key issue is how to extract from these movies physically relevant observables in the underlying quantum theory which hold to leading order in $1/N_c$.   The first step in doing so is to find quantities which can be identified clearly in the classical calculation (movie) which has physical meaning even in the quantum context. An example might be the fraction of the initial kinetic energy of the baryons which gets converted into energy in mesons (including the meson masses).  In any given classical solution of the Skyrme model this could be identified and similarly in any physical scattering process this could---at least in principal---be measured.   The natural way to quantify the physical observable in scattering is in terms of some sort of cross-section. Thus for example one could ask for the cross-section for collisions in which the fraction of initial energy is converted into mesons is greater than some fixed value, $f$.  This quantity is well-defined experimentally.  Quantum mechanically it can be obtained by summing over a large number of S-matrix elements (the number will scale with $N_c$ since kinematically the number of open channels that can contribute scales with $N_c$).  

The central point is that this quantity is also calculable classically and at large $N_c$; the classical result correctly accounts for the quantum results  up to $1/N_c$ corrections  (assuming that the model correctly describes the QCD physics). The cross-section is calculable classically in the following way: for any given classical solution with given impact parameter one can identify the fraction of the initial kinetic energy converted to meson field energy.  One can map out the region in impact parameter space for which this fraction is greater than $f$. The classical cross-section is simply the area in impact dimension space corresponding to such solutions.  Note that this cross-section is independent of $N_c$ in the large $N_c$ limit since the ``movie'' associated with the solution is independent of $N_c$.  Note moreover that an analogous procedure can be done to compute the cross-section for any property which is similarly independent of $N_c$. 
 
However, this procedure does not yet give the cross-section of physical interest.  Note that the classical calculation still depends on the variables associated with initial orientations of the hedgehogs, $A_1$ and $A_2$.  Thus, the cross-section associated with some property $\mathcal{P}$ computed classically is a function of $A_1$ and $A_2$: $\sigma^{\mathcal{P}} (A_1,A_2;p)$ where $p$ is the initial momentum and of order $N_c^1$.  In contrast the physical scattering observables are typically given in terms of the spin and flavor of the incident baryons with some spin quantization axis: $\sigma^{\mathcal{P}}_{m_s^1 \, m_i^1 \, m_s^2 \,  m_i^2}$. Since the directions of the spin projections are in principle arbitrary, the natural way to write the cross-section is as the expectation value of a quantum mechanical operator in spin-isospin space. The most general form consistent with isospin symmetry, time-reversal and parity is 
\begin{widetext}
\begin{align}
&\sigma^{\mathcal{P}}_{m_s^1 \, m_i^1 \, m_s^2 \,  m_i^2}(p) =  \langle m_s^1 \, m_i^1 \, m_s^2 \,  m_i^2| \hat{\sigma}^{\mathcal{P}}(p)  |m_s^1 \, m_i^1 \, m_s^2 \,  m_i^2 \rangle \; \; \; \; {\rm with}\label{sigmaform1}\\
&\hat{\sigma}^{\mathcal{P}}(p) \equiv  X_0(p) +  Y_0(p) \sigma_1 \cdot \sigma_2 +Z_0(p) (\sigma_1 \cdot \hat{n})(\sigma_2 \cdot \hat{n})  + \Big(X_1(p) + Y_1(p) \sigma_1 \cdot \sigma_2  + Z_1(p) (\sigma_1 \cdot \hat{n})(\sigma_2 \cdot \hat{n})\Big) \tau_1 \cdot \tau_2  \nonumber
\end{align}
\end{widetext}
where the hat in $\hat{\sigma}^{\mathcal{P}}(p)$ indicates that it is a quantum operator in spin-flavor space; in contrast, the hat in $\hat{n}$ indicates that it is a unit vector which is in the direction of of the scattering axis.  $p$ is the  momentum of the particle in the lab frame.   
Thus, there are in general six functions of the initial velocity (and possibly other variables associated with the observable) that fully characterize the cross-section associated with property ${\cal P}$: $X_0, X_1, Y_0, Y_1, Z_0$ and $Z_1$.

The critical issue is to relate  $\sigma^{\mathcal{P}} (A_1,A_2)(p)$, the classical cross-section parametrized by the orientation variables $A$, with  $X_0(p), X_1(p), Y_0(p), Y_1,(p), Z_0(p)$ and $Z_1(p)$. To do so,  it is sufficient to recall that the initial orientation variables $A$ are collective coordinates which are adiabatic and at large $N_c$ decouple from the other degrees of freedom in the problem.  Hence, once can promote these to quantum variables and requantize the motion associated with them.  The act of doing so converts classical configurations which break both spin and flavor into states with well defined spin and flavor quantum numbers.  This means that if the initial nucleons are in some spin-flavor state $|\psi\rangle$, then up to $1/N_c$ corrections, the cross-section  associated with ${\mathcal P}$ will be given by
\begin{align}
&\sigma^{\mathcal{P}}_{\psi}(p) \equiv \langle \psi |\hat{\sigma}^{\mathcal{P}}|\psi \rangle \\ & =\int dA_1 dA_2 \, |\psi (A_1,A_2)|^2  \, \sigma^{\mathcal{P}} (A_1,A_2;v)  \nonumber
\end{align} 
where the function $ \sigma^{\mathcal{P}} (A_1,A_2;p)$ is obtainable classically.  The collective wave-functions for states associated with well-defined spin and isospin are simply normalized Wigner $D$-matrices.  This implies that the leading order expression is given by
\begin{widetext}
\begin{equation}
\sigma^{\mathcal{P}}_{m_s^1 \, m_i^1 \, m_s^2 \,  m_i^2} =  \langle m_s^1 \, m_i^1 \, m_s^2 \,  m_i^2| \hat{\sigma}^{\mathcal{P}}  |m_s^1 \, m_i^1 \, m_s^2 \,  m_i^2 \rangle 
=  \frac{\int dA_1 dA_2 \, | D^{\frac{1}{2}}_{m_s^1 , m_i^1}(A_1)|^2 \,  | D^{\frac{1}{2}}_{m_s^2 , m_i^2}(A_2)|^2 \, \sigma^{\mathcal{P}} (A_1,A_2) }{\int dA_1 dA_2 \, | D^{\frac{1}{2}}_{m_s^1 , m_i^1}(A_1)|^2 \,  | D^{\frac{1}{2}}_{m_s^2 , m_i^2}(A_2)|^2} \; .\label{sigmaform2}
\end{equation}
\end{widetext}

The fact that the collective wave-functions are simply normalized Wigner $D$-matrices allows one to deduce a critical result: at leading order in $1/N_c$, $ \sigma^{\mathcal{P}}$ must be invariant under a simultaneous flip of spin and isospin for either of the incident nucleons.   This follows from a well-known property of Wigner $D$-matrices:
\begin{equation}
D^{j}_{m, m'}(A)=(-1)^{m-m'} \left( D^{j}_{-m, -m'} (A) \right)^* \, 
\label{D}\end{equation}
which implies that  $ \lvert D^{j}_{m, m'}(A) \rvert^2 = \lvert D^{j}_{-m,- m'}(A) \rvert^2 $.    Combining this with  Eq.~(\ref{sigmaform2}), implies that at leading order in the $1/N_c$ expansion
\begin{equation}
\begin{split}
\sigma^{\mathcal{P}}_{m_s^1 \, m_i^1 \, m_s^2 \,  m_i^2} (p)& = \sigma^{\mathcal{P}}_{-m_s^1 \, -m_i^1 \, m_s^2 \,  m_i^2}(p)\\&  =\sigma^{\mathcal{P}}_{m_s^1 \, m_i^1 \, -m_s^2 \,  -m_i^2}(p)  
\end{split}\label{cond1}
 \end{equation}
{\rm i.e.} a simultaneous flip of spin and isospin for either of the incident nucleons leaves the cross-section invariant.  On the other hand, in general $ \lvert D^{j}_{m, m'}(A) \rvert^2 \neq \lvert D^{j}_{-m, m'}(A) \rvert^2 $ and $ \lvert D^{j}_{m, m'}(A) \rvert^2 \neq \lvert D^{j}_{m, -m'}(A) \rvert^2 $.   Thus 
\begin{equation}
\begin{split}
\sigma^{\mathcal{P}}_{m_s^1 \, m_i^1 \, m_s^2 \,  m_i^2}(p) & \neq \sigma^{\mathcal{P}}_{-m_s^1 \, m_i^1 \, m_s^2 \,  m_i^2} (p)\\
\sigma^{\mathcal{P}}_{m_s^1 \, m_i^1 \, m_s^2 \,  m_i^2}(p) & \neq \sigma^{\mathcal{P}}_{m_s^1 \, -m_i^1 \, m_s^2 \,  m_i^2} (p)\\
\sigma^{\mathcal{P}}_{m_s^1 \, m_i^1 \, m_s^2 \,  m_i^2}(p) & \neq \sigma^{\mathcal{P}}_{m_s^1 \, m_i^1 \, -m_s^2 \,  m_i^2} (p)\\
\sigma^{\mathcal{P}}_{m_s^1 \, m_i^1 \, m_s^2 \,  m_i^2}(p) & \neq \sigma^{\mathcal{P}}_{m_s^1 \, m_i^1 \, m_s^2 \,  -m_i^2} (p)
\label{cond2}
\end{split}
\end{equation}
{{\it i.e.} a flip of only the spin or only  the isospin for one of the two nucleons alters the cross-section at leading order.

Comparing Eqs. (\ref{cond1}) and (\ref{cond2}) with Eq.~(\ref{sigmaform1}), it is apparent that of the six functions which parameterize the cross-section in general, $X_0, X_1, Y_0, Y_1, Z_0$ and $Z_1$. only three, $X_0$, $Y_1$ and $Z_1$ contribute at leading order in the $1/N_c$ approximation:
\begin{widetext}
\begin{equation}
\hat{\sigma}^{\mathcal{P}} =  \Big ( X_0(p) + \big ( Y_1(p) \sigma_1 \cdot \sigma_2  + Z_1(p) (\sigma_1 \cdot \hat{n})(\sigma_2 \cdot \hat{n}) \big ) \tau_1 \cdot \tau_2 \Big ) \times \left ( 1+ {\cal O}\left ( \frac{1}{N_c} \right) \right)
\label{result1} \end{equation}
\end{widetext}

Equation (\ref{result1}) has significant predictive power at least if the $1/N_c$ corrections are negligibly small.  Note that half of the possible terms allowed by rotation and isospin symmetries are absent in the leading order expression.  This means that certain quantities which could in principle differ will be identical  at large $N_c$.  

As an example, consider scattering where either the beam or the target (or both) are unpolarized.  Even if only the beam or the target is unpolarized it is clear from the general structure of  scattering in Eq.~(\ref{sigmaform1}) that polarization of the other particle is irrelevant. Since polarization is irrelevant, only $X_0$ and $X_1$ can contribute to scattering in Eq. (\ref{cond1}).  Of course this structure encodes isospin invariance, which implies that $\sigma^{\mathcal{P}}_{pp\, \rm unpolarized}(p) =\sigma^{\mathcal{P}}_{nn\, \rm unpolarized}(p)$.  However if $X_1 \neq 0$, as is generally expected in the absence of a symmetry reason for it to vanish, one sees that  $\sigma^{\mathcal{P}}_{pp\, \rm unpolarized}(p) \neq \sigma^{\mathcal{P}}_{np\, \rm unpolarized}(p)$.  However, as the large $N_c$ limit is approached,  a contracted spin-flavor symmetry emerges and  $X_1 \rightarrow 0$.  Thus for large $N_c$, Eq.~(\ref{result1}) implies that
\begin{widetext}
\begin{equation}
\sigma^{\mathcal{P}}_{pp\, \rm unpolarized}(p)  = \sigma^{\mathcal{P}}_{np\, \rm unpolarized}(p) \times \left ( 1+ {\cal O}\left ( \frac{1}{N_c} \right) \right)
\label{unpol} \end{equation} \end{widetext}
which is a prediction.   Once polarized beams and targets are considered a large number of similar predictions can be made\cite{cohen1}.  It should be clear that the predictive power is due to the emergent spin-flavor symmetry.  Note the analysis holds regardless of the parameters in the Skyrme Lagrangian and thus there are strong reasons to believe that it is a true model-independent prediction.

It is worth recalling that predictions such as the one in Eq.~(\ref{unpol}) are valid only in Witten kinematics which means that the incident momenta are taken to be of order $N_c$ and for cross-sections associated with properties ${\mathcal P}$ which can be computed in classical or mean-field treatments.  Recall also that ${\mathcal P}$ is typically associated with bulk properties involving energy or the flow of baryon number.  Thus for example, one can let ${\mathcal P}$ indicate the cross-section for scattering in which the baryons are deflected by more than some fixed angle (with any number of pions produced).  By differentiating with respect  to the angle one can obtain differential cross-sections.  However, it should be stressed that these differential cross-sections are {\it not} the usual differential cross-section for elastic scattering but rather a kind of semi-inclusive differential cross-section which has a fixed angle for the outgoing nucleons but includes any number of created mesons.

\section{Nucleon-Antinucleon Scattering}

 The problem of nucleon-antinucleon scattering was also considered in Witten's classic paper.  He again justified a time-dependent mean-field approach.  The central observation at the core of the present paper is that the spin-flavor dependence of certain classes of  nucleon-antinucleon scattering observables in Witten kinematics can be obtained in an analysis
 nearly identical to that of nucleon-nucleon scattering.  The  analysis of nucleon-antinucleon scattering applies to qualitatively the same types of observables, cross-sections associated with bulk properties involving energy or the flow of baryon number---{\it i.e.} things computable at the classical or mean-field level---in the Witten limit.  It also leads to the same type of spin-flavor dependence:
 \begin{widetext}
\begin{align}
&\sigma^{\mathcal{P}}_{m_s^{\rm N} \, m_i^{\rm N} \, m_s^{\overline{\rm N} }\,  m_i^{\overline{\rm N}} }(p) =  \langle m_s^{\rm N} \, m_i^{\rm N} \, m_s^{\overline{\rm N} }\,  m_i^{\overline{\rm N}} | \hat{\sigma}^{\mathcal{P}}(p)  |m_s^{\rm N} \, m_i^{\rm N} \, m_s^{\overline{\rm N} }\,  m_i^{\overline{\rm N}} \rangle  \; \; \; \;  {\rm with}\label{sigmafornnbar} \\
& \hat{\sigma}^{\mathcal{P}} (p) =  X_0(p) +  Y_0(p) \,  \sigma_{\rm N} \cdot \sigma_{\overline{\rm N}}   +Z_0(p) (\sigma_{\rm N} \cdot \hat{n})(\sigma_{\overline{\rm N}} \cdot \hat{n})  + \Big(X_1(p) + Y_1(p) \sigma_{\rm N} \cdot \sigma_{\overline{\rm N}}  + Z_1(p) (\sigma_{\rm N} \cdot \hat{n})(\sigma_{\overline{\rm N}} \cdot \hat{n})\Big) \tau_{\rm N} \cdot \tau_{\overline{\rm N}}  \nonumber\\
& {\rm where} \; \;  \;  \; X_0(p) \sim N_c^0 ,\; \; Y_1(p) \sim N_c^0 ,\; \; Z_1(p) \sim N_c^0     \; \; \; \;
{\rm and} \;  \; \; \;  X_1(p) \sim \frac{1}{N_c}, \; \; Y_0(p) \sim  \frac{1}{N_c}, \; \; Z_0(p) \sim  \frac{1}{N_c}  \nonumber 
\end{align}

\begin{center}
\begin{table}
\begin{center}
	\begin{tabular}{|c|c|}
	\hline
Relation number & Annihilation Cross-Section Relation\\
     \hline
     \hline
     1)&$\sigma^{\rm A}_{p \overline{n}\, \rm unpolarized}(p) = \sigma^{\rm A}_{p \overline{p} \, \rm unpolarized}(p) \times \left ( 1+ {\cal O}\left ( \frac{1}{N_c} \right) \right)$\\ \hline
      2)& $  \sigma^{\rm A \, L}_{p \overline{n}\, \uparrow \uparrow}(p)   =  \sigma^{\rm A \, L}_{p \overline{p}\, \uparrow \downarrow}(p)  \times \left ( 1+ {\cal O}\left ( \frac{1}{N_c} \right) \right)$\\ \hline
     3)& $  \sigma^{\rm A \, L}_{p \overline{n}\, \uparrow \downarrow}(p)   =  \sigma^{\rm A \, L}_{p \overline{p}\, \uparrow \uparrow}(p)  \times \left ( 1+ {\cal O}\left ( \frac{1}{N_c} \right) \right)$\\ \hline
      4)& $\frac{1}{2} \left (  \sigma^{\rm A \, L}_{p \overline{n}\, \uparrow \uparrow}(p)  + \sigma^{\rm A \, L}_{p \overline{p}\, \uparrow \uparrow}(p)  \right ) = \sigma^{\rm A}_{p \overline{n}\, \rm unpolarized}(p) \times \left ( 1+ {\cal O}\left ( \frac{1}{N_c} \right) \right)$\\ \hline
           5)& $\frac{1}{2} \left (  \sigma^{\rm A \, L}_{p \overline{n}\, \uparrow \uparrow}(p)  - \sigma^{\rm A \, L}_{p \overline{n}\, \uparrow \downarrow}(p)  \right ) = - \frac{1}{2} \left (  \sigma^{\rm A \, L}_{p \overline{p}\, \uparrow \uparrow}(p)  - \sigma^{\rm A \, L}_{p \overline{p}\, \uparrow \downarrow}(p)  \right )\times \left ( 1+ {\cal O}\left ( \frac{1}{N_c} \right) \right)$\\
     \hline
 6)& $  \sigma^{\rm A \, T}_{p \overline{n}\, \uparrow \uparrow}(p)   =  \sigma^{\rm A \, T}_{p \overline{p}\, \uparrow \downarrow}(p)  \times \left ( 1+ {\cal O}\left ( \frac{1}{N_c} \right) \right)$\\ \hline
     7)& $  \sigma^{{\rm A} \, T }_{p \overline{n}\, \uparrow \downarrow}(p)   =  \sigma^{\rm A \, T}_{p \overline{p}\, \uparrow \uparrow}(p)  \times \left ( 1+ {\cal O}\left ( \frac{1}{N_c} \right) \right)$\\ \hline
      8)& $\frac{1}{2} \left (  \sigma^{\rm A \, T}_{p \overline{n}\, \uparrow \uparrow}(p)  + \sigma^{\rm A \, T}_{p \overline{p}\, \uparrow \uparrow}(p)  \right ) = \sigma^{\rm A}_{p \overline{n}\, \rm unpolarized}(p) \times \left ( 1+ {\cal O}\left ( \frac{1}{N_c} \right) \right)$\\ \hline
           9)& $\frac{1}{2} \left (  \sigma^{\rm A \, T}_{p \overline{n}\, \uparrow \uparrow}(p)  - \sigma^{\rm A \, T}_{p \overline{n}\, \uparrow \downarrow}(p)  \right ) = - \frac{1}{2} \left (  \sigma^{\rm A \, T}_{p \overline{p}\, \uparrow \uparrow}(p)  - \sigma^{\rm A \, T}_{p \overline{p}\, \uparrow \downarrow}(p)  \right )\times \left ( 1+ {\cal O}\left ( \frac{1}{N_c} \right) \right)$\\
     \hline
\end{tabular}
\end{center}
\caption{Relations between the total annihilation cross-section for nucleon-antinucleon scattering  in various spin and flavor channels that hold for sufficiently large $N_c$.  The superscript $A$ in $\sigma^{{\rm A}}$ indicates that the cross-section is for the total annihilation cross-section.  The superscripts  L and T stand for longitudinal and transverse respectively and indicate the polarization axis; longitudinal corresponds to spins polarized along the beam axis and transverse is for spins polarized perpendicular to it.  Thus, $\sigma^{\rm A \, L}_{p \overline{n}\, \uparrow \uparrow}$ corresponds to a proton and antineutron with both spins $+\frac{1}{2}$ quantized along the positive beam direction.  The subscript ``unpolarized'' corresponds to situations when either the nucleon or the antinucleon (or both) are unpolarized.}
\label{table} 
\end{table}
\end{center}
\end{widetext}

The logic underlying this is analogous to the  nucleon-nucleon case.  Once again, for the purpose of identifying model-independent relations, models based on topological solitons serve as a surrogate for the full problem.  Again one  begins the analysis by imagining access to an arbitrarily  large set of solutions of the classical (or mean-field) scattering processes associated with a given initial velocity.    In  this case, these classical solutions all involve initial conditions of two widely separated rotated hedgehog baryons, one with winding number one (a baryon) and the other with winding number negative one (an antibaryon).  They move towards each other offset by an impact parameter ${\bf b}$, a two dimensional vector.  It is natural to restrict one's attention to situations in which the baryon is initially at rest and the antibaryon moving towards it, as this is the typical experimental set up.   Thus,  as in the case of nucleon-nucleon scattering, the initial conditions depend on the momentum, the scattering axis and 8 more collective variables---2 for the components of the impact parameter vector and 6 to specify the orientation of the two hedgehogs.  Again for fixed values of the orientation variables, one can map out which values of the  impact  lead to classical solutions satisfying property ${\mathcal P}$; the area in impact parameter space defines the cross-section $\sigma^{\mathcal{P}} (A_1,A_2;p)$.  Again  the orientation degrees of freedom are requantized.  Finally, a comparison of the most general form for the cross-section to the form obtained from taking matrix elements of  $\sigma^{\mathcal{P}} (A_1,A_2;p)$ using the requantized $A$ variables and exploiting Eq.~(\ref{D}) leads to Eq.~(\ref{sigmafornnbar})

\section{Total Annihilation Cross-Section}

As noted in the introduction, the type of observables in nucleon-nucleon scattering for which the spin-flavor dependence is determined at large $N_c$ are not those which experimentalists  analyze and report.  The analogous observables in nucleon-antinucleon scattering are also typically not analyzed in experiments.  However, nucleon-antinucleon scattering has one observable that has no analog in nucleon-nucleon scattering, is in the class of observables for which the large $N_c$ analysis applies and {\it is} commonly analyzed in experiment: the total annihilation cross-section.

It is easy to see that the total annihilation cross-section is in the class of observables for which the analysis applies. In effect, this means that it can be calculated classically for any given initial classical configuration parameterized by the hedgehog, orientation angles, the impact parameter vector and the velocity.  Starting with an initial configuration in the Skyrme model, one can follow the field configuration forward in time.  The baryon density $\rho_B$ is given by the topological winding number density, which is fixed by the field configuration.  Initially the baryon density distribution consists of two well separated blobs, one with a baryon number one and the other with baryon number of negative one, which are traveling towards each other, off-set by the impact parameter.  As time goes forward, the distribution of baryon density may get complicated as the baryon and antibaryon interact.  However, at sufficiently long times after the interaction begins, things will necessarily simplify: as $t \rightarrow \infty$, either $\rho_B$, the baryon density, will approach  zero everywhere (corresponding to annihilation) or it will be in the form of two lumps, one with integrated baryon number one and with baryon number of negative one moving away from each other with a velocity whose magnitude is less than the initial velocity (corresponding to an inelastic scattering process in which energy is lost to meson emission).  Which of these two outcomes occurs is completely determined by the orientation angles, the impact parameter vector and the momentum---this is a necessary consequence of the calculation being classical and hence deterministic.  The classical annihilation cross-section for fixed $p$, $A_1$, and $A_2$ is simply the area in impact parameter space of initial configurations for which the $\rho_B(\vec{x},t) \rightarrow 0$  for all $\vec{x}$ as $t \rightarrow \infty$.

The upshot of this is that the spin-isospin dependence of the total annihilation cross-section at large $N_c$ is given by  Eq.~(\ref{sigmafornnbar}).  To the extent that the subleading terms in $1/N_c$ are negligible, it means that all total annihilation cross-sections are expressible in terms of just  three functions of the initial velocity.  This enables one to relate the total annihilation cross-section in many different spin-isospin channels up to corrections of relative order $1/N_c$.  A number of these relations are given in Table \ref{table}.  All of the relations in Table \ref{table} are true predictions of large $N_c$ QCD; that is, these relations depend on more than mere isospin and rotational invariance.   A quick glance at the table makes clear that relations 2 and 6, 3 and 7, 4 and 8, and 5 and 9 have identical forms. The only difference is that one relation applies for longitudinal polarization and the other applies for transverse.  However, despite the formal structures, they are distinct predictions--in general, the annihilation cross-section for transversely polarized nucleons and antinucleons differ, even at large $N_c$.

The list of relations in Table \ref{table} is not complete.  Clearly one can exploit rotational invariance or isospin invariance to generate new relations from those in the table.  Thus, if one simultaneously  switches all protons with neutrons and antiprotons with antineutrons, all the relations remain true; similarly if one swaps all spin up states and spin down for both particles simultaneously, all of the relations remain true.   One can  also generate new relations by taking linear combinations of the existing relations.  Finally one can also generate new relations from Eq.~(\ref{sigmafornnbar}) for situations where the spin polarization directions are neither longitudinal nor transverse but at any given angle to the beam and for cases where the polarizations of the nucleon and antinucleon are quantized along different axes.  

One of the principle reasons for the focus on the total annihilation cross-section---apart from the fact that it is an intrinsically interesting quantity---is that it is one which can be studied experimentally.  Unfortunately, to the best of our knowledge, experiments measuring annihilation cross-section with polarized antineutron beams have not been done.  This is hardly surprising since such an experiment would be immensely challenging.  This means that most of the relations in Table \ref{table} cannot be tested empirically.  Fortunately,  relation 1) does not depend on polarized beams and ought to be testable.  It predicts that the total annihilation cross-section for proton-antiproton and proton-antineutron reactions should be the same up to $1/N_c$ corrections when the system is in Witten kinematics.

Before confronting relation 1) with experimental data, it is important to clarify the nature of Witten kinematics, {\it i.e.} the requirement that  $p$ is of order $N_c$.  Formally this means that a nucleon-nucleon or nucleon-antinucleon cross-section associated with property ${\mathcal P}$ in spin-isospin channel $c$ satisfies 
\begin{equation}
\begin{split}
& \sigma^{\mathcal{P}}_c(p) =f^{\mathcal P}_c(\tilde{p}) \times \left ( 1+ {\cal O}\left ( \frac{1}{N_c} \right ) \right )  \\
&{\rm with} \; \; \tilde{p} \equiv p/N_c
\end{split}
\end{equation}
where $f^{\mathcal P}_c(\tilde{p})$ is independent of $N_c$.  It is important to note that the coefficient in front of the relative order  $\left ( \frac{1}{N_c} \right )$ correction can depend both on the ${\mathcal P}$, the property defining the cross-section,  and $\tilde{p} \equiv p/N_c$.   It is possible that this coefficient grows with decreasing $\tilde{p}$,  and indeed, that for certain choices of  ${\mathcal P}$, the coefficient diverges as $\tilde{p} \rightarrow 0$ indicating a breakdown of the $1/N_c$ expansion at $\tilde{p}=0$.  This can happen for the following reason:  at any $N_c$, including arbitrarily large ones, $\tilde{p}=0$ corresponds to $p=0$ while for any non-zero value of $\tilde{p}$ at sufficiently large $N_c$, $p$ is much larger than characteristic hadronic scales which such mesons masses and the size of the nucleon which might be expected to control the onset of semi-classical/mean-field regime on which the large $N_c$ analysis is based.   

Given this, there is a concern that at sufficiently low $\tilde{p}$ for any given value of $N_c$,  the system might be outside the regime of validity of the $1/N_c$ analysis and the spin-flavor predictions based on it cease to be valid.    In the case of nucleon-nucleon scattering, it is clear that this happens essentially regardless of the property, ${\mathcal P}$.  Note that as $p \rightarrow 0$, the scattering becomes purely elastic s-wave scattering and involves a single quantum mechanical channel.  Since this is fundamentally quantum in nature, one expects that any semi-classical analysis  should break down at low enough momenta.  Empirically, it was seen in ref.~\cite{cohen4} that the large $N_c$ predictions of the spin-flavor dependence failed when used in the elastic scattering of nucleon-nucleon scattering.   If a similar breakdown of the validity of the large $N_c$ analysis  occurs for nucleon-antinucleon scattering at low momentum, it will be difficult to assess empirically  the predictive power of the large $N_c$ approach in nucleon-antinucleon annihilation, since  the antineutron-proton annihilation data is at rather low beam momenta---below 500 MeV.

Fortunately, low momentum nucleon-antinucleon annihilation differs from low momentum nucleon-nucleon scattering in a fundamental way.  While nucleon-nucleon scattering at very low momentum involves only a single quantum channel (the elastic s-wave channel), nucleon-antinucleon annihilation always involves many channels---even as the incident momentum goes to zero.  The decay can go into two mesons in s-wave, three mesons in various angular momentum and isospin combinations, 4 mesons in multiple combinations, etc.  As the $N_c$ limit is approached, the number of such channels grows rapidly as a function of $N_c$.    Since the crux of how a classical result emerges from a quantum scattering problem is the contributions from many quantum channels which behave similarly, and the classical regime was key to the derivation of the spin-flavor relations for scattering, it is plausible that what is driving the validity of the spin-flavor dependence for scattering is the presence of many kinematically available quantum channels.    If this is the case, then one expects the relations in Table \ref{table} to hold at large $N_c$ for arbitrarily small values of $\tilde{p}$.  While this is not guaranteed to be correct, it is plausible and motivates a comparisons  of relation 1) with the available data, even though the data is at low momentum.

\begin{figure}[t]
\includegraphics[width=.45\textwidth]{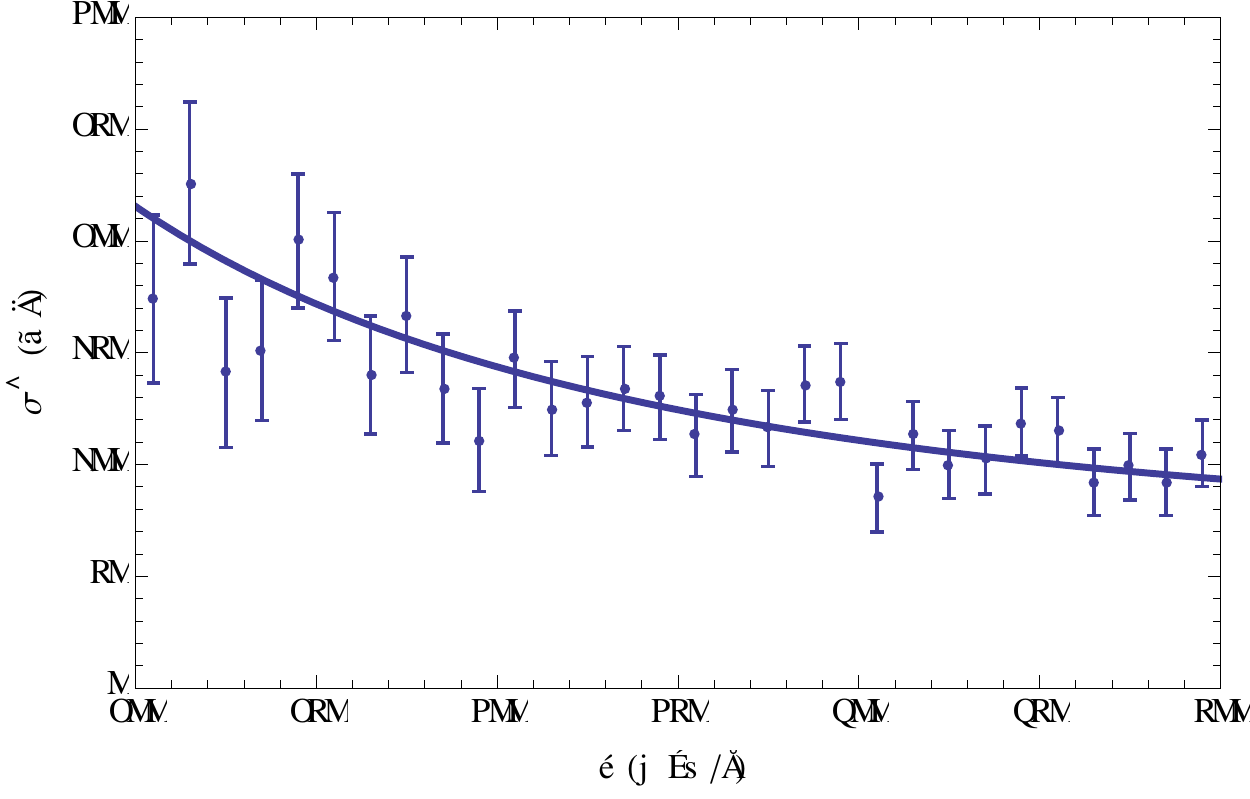}
\caption{The data points represent the total annihilation cross-section for unpolarized antineutrons incident on protons with various beam momentum; the data comes from ref. \cite{pn} and the errors include both systematic and statistical errors added in quadrature.  The solid curve is a phenomenological fit to the total annihilation cross-section for unpolarized antiprotons incident on protons  as a function of beam  momentum from ref.~\cite{pp}.  This curve accurately fits all of the data points in this range to within a few percent. }
\label{fig:f1}
\end{figure}

In Fig.~\ref{fig:f1}, data from ref.~\cite{pn} for annihilation cross-section of an unpolarized antineutron beam on a proton target is given as a function of beam momentum.   To the extent that relation 1) of Table \ref{table} applies, the antiproton-proton annihilation cross-sections should be the same up to creations of relative order $1/N_c$.  For ease of comparison, rather than giving individual data points for the antiproton-proton annihilation cross-section, a fit to this data taken from ref.~\cite{pp} is given. The curve is entirely phenomenological in nature and is not based on any underlying theory.  However, the quality of the fit is extremely good.   The key point is that the quality of this fit is sufficiently good that for all data points, the disagreement between the fit and the data is much smaller  than the quoted error bars for any of the antineutron-proton data points.   It thus seems that the fit is a useful basis for comparison with the antineuton-proton data.  In practice, the antiproton-proton data points at the lower end of this range are accurately fit to within a few percent by the curve and most of the points are fit to better than a percent.   

The comparison between the fitted antiproton-proton data and the antineutron-proton data is rather striking.  By eye, it seems that the expectation based relation 1) holds.  Indeed, the fit {\it looks} as though it could have been based on the antineutron-proton data rather than the antiproton-proton data.  One way to quantify this: the  $\chi^2$ per degree of freedom of the anti-neutron data taking the curve fit to antiproton-proton as the theoretical prediction is less than unity.  This is remarkable, in that even if relation 1) is valid, it only expected to hold to order $1/N_c$ which for the real world is 1/3.  Thus, given given the quality of the data, it seems quite safe to conclude that empirically relation 1) does hold.

Of course, the fact that the unpolarized  antiproton-proton and antineutron-proton total annihilation cross-sections are so similar to what is expected at large $N_c$ does not necessarily indicate that this is due to the viability of large $N_c$  physics for scattering  at $N_c=3$.  It is quite possible that the agreement with the large $N_c$ prediction is accidental and stems from some other cause.  The case would be strengthened significantly if the data extended to significantly higher momenta (say up 2 GeV) where the question of whether the momentum is high enough for  system  to be in the regime of Witten kinematics does not arise.  More compelling would be data with polarized beams acting on polarized targets  using both antineutron and antiproton beams.  This would allow tests for more than just relation 1). However, despite the limitations of the data, it is nevertheless encouraging that the data we do have is consistent with the large $N_c$ expectations.

\end{document}